\documentclass[aps,prd,twocolumn,nofootinbib,superscriptaddress]{revtex4-1}
\usepackage{amsmath,amssymb}
\usepackage{graphicx}
\usepackage{hyperref}
\hypersetup{colorlinks=true,allcolors=blue}

\begin{document}

\title{Numerical tests of 5D APS $\eta$-invariant phases for 4D chiral fermions}
\author{Tatsuhiro Misumi}
\email{misumi@phys.kindai.ac.jp}
\affiliation{Department of Physics, Kindai University, Higashi-Osaka, Osaka 577-8502, Japan}

\begin{abstract}
We calculate the phase of a 4D chiral fermion determinant from the 5D Atiyah--Patodi--Singer (APS) $\eta$-invariant. We study the anomaly-free $SU(5)$ representation $\mathbf{10}\oplus\overline{\mathbf5}$ on a four-torus, using gauge fields independent of one coordinate. For these backgrounds, the 4D phase can be calculated independently from a 3D Dirac spectrum. For an Abelian background embedded in $SU(5)$, two different 5D bulk extensions reproduce the 4D phase within $0.03\%$ in the tested heat-kernel extrapolations. One extension changes the spatial gauge field along the fifth direction and requires a coupled 5D spectral calculation. An independent single-charge check reproduces the Chern--Simons anomaly coefficient within $0.3\%$ using quadratic extrapolations. We also calculate 5D APS phases for non-Abelian backgrounds in the $SU(4)\times U(1)$ subgroup. For a background generating the full $\mathfrak{su}(4)$ algebra, the 5D phase with a fixed spatial field agrees with the independent 4D phase within $0.5\%$ over the six heat-kernel parameters examined. We further study a varying-field 5D extension for an embedded $SU(2)$ background. These results show that the 5D APS calculation reproduces a nonzero chiral phase after anomaly cancellation.
\end{abstract}

\maketitle

%%%%%%%%%%%%%%%%%%%%%%%%%%%%%%%%%%%%%%%%%%%%%%%%%%%%%%%%%%%%%%%%%
\section{Introduction}
\label{sec:intro}

A nonperturbative formulation of chiral gauge theories remains a challenge in lattice field theory. The Nielsen--Ninomiya theorem shows the difficulty of describing a single Weyl fermion with a local lattice Dirac operator under its standard assumptions~\cite{Nielsen:1980rz}. Domain-wall and overlap formulations provide ways to obtain chiral fermions~\cite{Kaplan:1992bt,Shamir:1993zy,Narayanan:1993sk,Narayanan:1994gw,Neuberger:1997fp}. The Ginsparg--Wilson relation provides a formulation of exact lattice chiral symmetry~\cite{Ginsparg:1981bj,Luscher:1998pqa}.

The fermion measure must also respect gauge invariance. Gauge-invariant measures have been constructed for anomaly-free Abelian theories, and their extension to non-Abelian theories has been studied extensively~\cite{Luscher:1998du,Luscher:1999un,Luscher:2000zd,Kikukawa:2001mw,Kadoh:2003ii,Kadoh:2004uu,Kadoh:2007wz,Kadoh:2007xb}. Other approaches use mirror-fermion decoupling, gauge-field flow, or higher-dimensional regularization~\cite{Poppitz:2007tu,Giedt:2007qg,Poppitz:2010at,Chen:2012di,Grabowska:2015qpk,Grabowska:2016bis,Fukaya:2016ofi,Kaplan:2023pxd,Kaplan:2023pvd,Kaplan:2024ezz,Clancy:2024bjb,Sen:2026pnm}. Recent studies have explored alternative formulations and constructions of lattice chiral theories~\cite{Craig:2026soq,Thorngren:2026ydw,Yonekura:2026local}.

Another approach starts with lattice gauge links, constructs a smooth continuum gauge field, and couples continuum Weyl fermions to that field. Continuum gauge fields constructed from lattice links have been considered for many years~\cite{Gockeler:1992ig,Gockeler:1992bi,tHooft:1994jaf,Hsu:1995qm,Kronfeld:1995ws,Hernandez:1995jc,Bodwin:1995rd,Hernandez:1996rd}. This approach has been studied numerically in two-dimensional chiral models~\cite{Montvay:1995qb,Gattringer:1995du,Gattringer:1996ke}.

A prescription in which the magnitude of the Weyl determinant is regularized by Pauli--Villars (PV) fields and its phase is obtained from a higher-dimensional $\eta$-invariant has recently been proposed~\cite{Hsu:2026gmi,Witten:2026ezr}. Conditions under which the relative weights of this construction and the overlap formulation agree in the continuum limit, after a local counterterm is fixed, have also been established~\cite{Craig:2026soq,Liu:2026matching}. The $\eta$-invariant is a regularized measure of the spectral asymmetry of a Dirac operator. Its relation to chiral determinants is described by the Atiyah--Patodi--Singer (APS) index theorem and subsequent work on determinant lines and anomaly inflow~\cite{Atiyah:1975jf,Alvarez-Gaume:1984zst,Dai:1994kq,Witten:2019bou}. Earlier studies related $\eta$-invariants to lattice and domain-wall fermions~\cite{Kaplan:1995pe,Tatsumi:1999jlf,Fukaya:1710.03379,Fukaya:1910.01987,Fukaya:1910.09675}. In these approaches, the $\eta$-invariant also provides a way to describe the APS index with domain-wall Dirac operators. It was pointed out that independence of the five-dimensional interpolation is not guaranteed at finite lattice spacing~\cite{Tatsumi:1999jlf}. This makes a numerical comparison of different bulk extensions particularly useful.

In our previous paper~\cite{Misumi:2026contf}, we calculated the PV magnitude and APS phase in the two-dimensional $3450$ model. The exact torus determinant provided a reference for the full complex fermion weight, and we tested the dependence on the bulk extension. Related two-dimensional chiral theories have also been studied with lattice symmetry, bosonization, and other nonperturbative methods~\cite{Berkowitz:2023pnz,Morikawa:2024zyd,Onogi:2025tev,Lootens:2026sim}. In 4D, an exact reference of this kind is harder to obtain because the chiral determinant is generally not known. The imaginary part of the 4D chiral $U(1)$ effective action was studied by coupling fermions on a finer lattice to interpolated gauge fields and taking the finer-lattice limit~\cite{Bornyakov:1999be,Bornyakov:9909135}. Here we calculate the phase directly from a 5D APS spectrum, following Refs.~\cite{Hsu:2026gmi,Witten:2026ezr}, and compare it with an independent 4D spectral calculation. In particular, we test the nonzero phase that remains when the local gauge anomaly cancels.

We consider the anomaly-free $SU(5)$ representation $\mathbf{10}\oplus\overline{\mathbf5}$. We choose four-dimensional gauge fields independent of one torus coordinate, with a constant gauge component along that direction that commutes with the spatial fields. The Fourier modes in the chosen direction can be summed analytically, reducing the 4D determinant phase to a convergent sum over 3D Dirac eigenvalues. The local anomalous terms cancel among the Weyl fermions, while a finite, nonzero phase survives. This gives an independent reference for the 5D APS calculation.

We show the agreement of the 5D APS phase with this independent 4D result for two smooth bulk extensions of the same Abelian boundary fields. The first extension keeps the spatial field fixed. In the second, the field changes inside the 5D bulk, so the calculation must retain the coupled spatial modes. We also calculate 5D APS phases for non-Abelian $SU(4)$ fields and find agreement with the independent 4D values. The non-Abelian tests include both a field generating the full $\mathfrak{su}(4)$ algebra and a varying bulk field in an embedded $SU(2)$ subgroup.

We use specified smooth continuum gauge fields, so that the comparison isolates the fermion phase from the reconstruction of continuum fields from lattice links. Section~\ref{sec:theory} defines the model, the independent 4D phase, and the 5D APS calculation. Sections~\ref{sec:abelian} and \ref{sec:nonabelian} present the Abelian and non-Abelian results, respectively. We summarize the results in Sec.~\ref{sec:summary}, with numerical details in the appendices.

%%%%%%%%%%%%%%%%%%%%%%%%%%%%%%%%%%%%%%%%%%%%%%%%%%%%%%%%%%%%%%%%%%%%%%%
\section{Model and analytic formulation}
\label{sec:theory}

%%%%%%%%%%%%%%%%%%%%%%%%%%
\subsection{Gauge fields and anomaly cancellation}

We work on a Euclidean four-torus with unit periods and coordinates $(x,y,z,w)$. We impose antiperiodic fermion boundary conditions around all four cycles. The chiral fermions belong to the anomaly-free $SU(5)$ representation
\begin{equation}
\mathcal R=\mathbf{10}\oplus\overline{\mathbf5}.
\label{eq:representation}
\end{equation}
We use Hermitian gauge potentials. In the fundamental representation, the covariant momentum is $-i\partial_\mu+A_\mu$, and the field strength is $F_{\mu\nu}=\partial_\mu A_\nu-\partial_\nu A_\mu+i[A_\mu,A_\nu]$. For the following backgrounds, it is convenient to introduce
\begin{equation}
Y=\operatorname{diag}(1,1,1,1,-4).
\label{eq:Y}
\end{equation}
Under $SU(4)\times U(1)\subset SU(5)$, the representation decomposes as
\begin{align}
\mathbf{10}&=\mathbf6_{2}\oplus\mathbf4_{-3},\nonumber\\
\overline{\mathbf5}&=\overline{\mathbf4}_{-1}\oplus\mathbf1_{4}.
\label{eq:branch}
\end{align}
The subscripts denote the eigenvalues of $Y$. The four sectors have charges $q=2,-3,-1,4$ and dimensions $n_q=6,4,4,1$, respectively. Together they give 15 left-handed Weyl components. In particular,
\begin{equation}
\sum_q n_q q=0,\qquad \sum_q n_q q^3=0.
\label{eq:cancel}
\end{equation}
The $SU(4)^3$ anomaly cancels between the fundamental $\mathbf4$ and the antifundamental $\overline{\mathbf4}$, whose cubic anomaly coefficients have opposite signs. The $\mathbf6$ is a real representation and has no cubic gauge anomaly. The mixed $SU(4)^2U(1)$ anomaly involves the quadratic index $T(R)$ of a representation $R$. It is defined by $\operatorname{tr}_R(t_R^at_R^b)=T(R)\delta^{ab}$, where $t_R^a$ are the $SU(4)$ generators in $R$, with $T(\mathbf4)=1/2$ and $T(\mathbf6)=1$ for $\mathbf6=\wedge^2\mathbf4$. The mixed anomaly coefficient is
\begin{equation}
2T(\mathbf6)-3T(\mathbf4)-T(\overline{\mathbf4})=0.
\label{eq:mixed}
\end{equation}
These identities are inherited from the anomaly-free $SU(5)$ representation.

We take the gauge potential in the form
\begin{equation}
\begin{split}
A_y&=hY,\\
A_j&=a_j(x,z,w)Y+
\begin{pmatrix}B_j(x,z,w)&0\\0&0\end{pmatrix},
\quad j=x,z,w,
\end{split}
\label{eq:background_general}
\end{equation}
where $a_j$ are real Abelian gauge potentials and $B_j$ are traceless Hermitian $4\times4$ matrices. All fields are independent of $y$, and $h$ fixes the Wilson line $\exp(ihY)$ around the $y$ cycle. As $Y$ commutes with $B_j$,
\begin{equation}
F_{yj}=0,\qquad \operatorname{tr}F\wedge F=0.
\label{eq:FwedgeF}
\end{equation}
Here $\operatorname{tr}$ denotes the trace in the fundamental representation. The second relation holds because the remaining curvature has components only along $x$, $z$, and $w$. It does not require the spatial gauge fields to commute.

For the Abelian test, $B_j=0$ and
\begin{equation}
 a_x=u\cos(2\pi kz),\quad a_z=0,\quad
 a_w=v\sin(2\pi kz).
\label{eq:abelian_background}
\end{equation}
Here $u$ and $v$ are real amplitudes and $k$ is an integer spatial wave number. Unless stated otherwise, we set
\begin{equation}
 (u,v,h,k)=(0.4,0.6,0.34,1).
\label{eq:parameters}
\end{equation}
All backgrounds below are smooth and globally defined on $T^4$, with vanishing topological charge $Q=(8\pi^2)^{-1}\int_{T^4}\operatorname{tr}(F\wedge F)=0$. The Abelian and non-Abelian backgrounds have no fermion zero modes. Below we verify the absence of spatial spectral crossings using the free gap and operator-norm bounds. We work in the $Q=0$ sector, in contrast to the nonzero-flux tests in Ref.~\cite{Misumi:2026contf}.

%%%%%%%%%%%%%%%%%%%%%%%%%%%%
\subsection{The 4D determinant from the 3D spectrum}
\label{sec:benchmark}

%%%%
\subsubsection{Compact-direction reduction}

%We compare two determinants with the same spatial gauge field but different values of the constant $y$ component, $h$ and $0$. Their ratio removes the overall normalization. For an $SU(4)$ representation $R$ with $U(1)$ charge $q$, define the Hermitian three-dimensional Dirac operator
We compare two determinants with the same spatial gauge field but different values of the constant $y$ component, $h$ and $0$.
Their ratio removes the overall normalization. For an $SU(4)$ representation $R$ with $U(1)$ charge $q$, where $(R,q)$ labels one of the four sectors in Eq.~(\ref{eq:branch}), we define the Hermitian three-dimensional Dirac operator
\begin{equation}
\begin{aligned}
H_{3,R}={}&\sigma_x(-i\partial_x+qa_x+B_x^R)\\
&+\sigma_z(-i\partial_z+qa_z+B_z^R)\\
&+\sigma_w(-i\partial_w+qa_w+B_w^R).
\end{aligned}
\label{eq:H3}
\end{equation}
Here $(\sigma_x,\sigma_z,\sigma_w)=(\sigma_1,\sigma_2,\sigma_3)$ are the Pauli matrices, and $B_j^R$ is the matrix representing the $SU(4)$ connection in $R$. Since $q$ is fixed by $R$ in Eq.~(\ref{eq:branch}), we suppress it in the notation $H_{3,R}$. We also write $H_3$ when the sector is understood. For the chirality used here, the 4D Weyl operator is
\begin{equation}
D_{+,R,q}=H_{3,R}+i(-i\partial_y+qh).
\label{eq:weyl}
\end{equation}
The chosen chirality fixes the sign of the phase. Reversing it complex conjugates the determinant.

Let $\lambda_{R,\alpha}\in\mathbb R$ denote the eigenvalues of $H_{3,R}$, with $\alpha$ labeling the spatial eigenstates, including their color components. The antiperiodic Fourier momentum in the $y$ direction is $2\pi(n+1/2)$. Denoting the determinant of Eq.~(\ref{eq:weyl}) by $W_{R,q}(h)$, the product over the $y$ modes gives
\begin{align}
\frac{W_{R,q}(h)}{W_{R,q}(0)}
&=\prod_\alpha \frac{\cosh[(\lambda_{R,\alpha}+iqh)/2]}{\cosh(\lambda_{R,\alpha}/2)},\nonumber\\
&=\prod_\alpha \left[\cos\frac{qh}{2}
+i\sin\frac{qh}{2}\tanh\frac{\lambda_{R,\alpha}}{2}\right].
\label{eq:kk}
\end{align}
The product over the $y$ modes is evaluated by symmetric pairing, as described in Appendix~\ref{app:product}. The phase of the remaining spatial product is defined using the $\eta$-invariant below. Related hyperbolic-function formulas occur in earlier calculations of parity-odd fermion determinants with a compact direction~\cite{Fosco:1997nonabelian}. Here Eq.~(\ref{eq:kk}) provides a numerical reference for the 5D APS result.

For a spatial eigenvalue $\lambda=\lambda_{R,\alpha}$, let $\vartheta_\lambda(qh)$ be the phase of the corresponding factor in Eq.~(\ref{eq:kk}), chosen continuously from $\vartheta_\lambda(0)=0$. For $|qh|<\pi$, it is
\begin{equation}
\vartheta_\lambda(qh)
=\frac12\int_0^{qh}dt\,
\frac{\sinh\lambda}{\cosh\lambda+\cos t},
\label{eq:single_phase}
\end{equation}
which is equivalent on this branch to $\arctan[\tan(qh/2)\tanh(\lambda/2)]$. We define $\Theta_{R,q}(h)$ as the continuous phase of the determinant ratio $W_{R,q}(h)/W_{R,q}(0)$, with $\Theta_{R,q}(0)=0$. Separating the large-$|\lambda|$ behavior gives an $\eta$-invariant term and an exponentially convergent sum,
\begin{align}
\Theta_{R,q}(h)={}&\frac{qh}{2}\eta(H_{3,R})\nonumber\\
&+\sum_\alpha\left[\vartheta_{\lambda_{R,\alpha}}(qh)
-\frac{qh}{2}\operatorname{sgn}(\lambda_{R,\alpha})\right].
\label{eq:eta_finite}
\end{align}
Here $\eta(H_{3,R})$ is obtained by analytically continuing $\eta(H,\nu)=\sum_{\lambda\ne0}\operatorname{sgn}(\lambda)|\lambda|^{-\nu}$ to $\nu=0$ and setting $H=H_{3,R}$. Each summand in the second line decreases exponentially with $|\lambda|$. In our backgrounds the spatial spectrum has a gap and the chosen values satisfy $|qh|<\pi$ for every charge.

The compact-direction product also gives the squared magnitude of each factor in Eq.~(\ref{eq:kk}),
\begin{equation}
\left|\frac{\cosh[(\lambda+iqh)/2]}{\cosh(\lambda/2)}\right|^2
=\frac{\cosh\lambda+\cos(qh)}{\cosh\lambda+1}.
\label{eq:magnitude}
\end{equation}
The product of these factors gives the squared magnitude of the determinant ratio.

%%%%
\subsubsection{Local and finite parts of the phase}

For a single charge-$q$ Weyl component in the Abelian background, we write $H_{3,q}=\sum_{j=x,z,w}\sigma_j(-i\partial_j+qa_j)$ and denote its free-field operator by $H_{3,q}^{(0)}$. We also write $a=a_j\,dx^j$ for the spatial Abelian gauge one-form. The change in the 3D $\eta$-invariant is fixed by the local Chern--Simons term, as follows from the parity anomaly~\cite{Alvarez-Gaume:1984zst,Dai:1994kq}. With no eigenvalue crossing and the orientation used here,
\begin{equation}
\eta(H_{3,q})-\eta(H_{3,q}^{(0)})
=\frac{q^2}{4\pi^2}\int_{T^3}a\wedge da
=\frac{q^2kuv}{2\pi}.
\label{eq:eta_abelian}
\end{equation}
Combining the factor $qh/2$ in Eq.~(\ref{eq:eta_finite}) with the $q^2$ dependence in Eq.~(\ref{eq:eta_abelian}), the $\eta$ term in the four-dimensional phase is proportional to $q^3$. We denote this local contribution to the phase of one charge-$q$ Weyl component by
$\Theta_{q,\eta}=(qh/2)[\eta(H_{3,q})-\eta(H_{3,q}^{(0)})]$, giving
\begin{equation}
\Theta_{q,\eta}=\frac{q^3kuvh}{4\pi}.
\label{eq:abelian_cs_phase}
\end{equation}
By Eq.~(\ref{eq:cancel}), it cancels in the complete chiral multiplet. Writing $W(h)=\prod_{R,q}W_{R,q}(h)$ for the full chiral determinant, we denote the continuous phase of $W(h)/W(0)$ by $\Theta_{4D}$. It is then
\begin{equation}
\Theta_{4D}=\sum_{R,q} \sum_\alpha
\left[\vartheta_{\lambda_{R,\alpha}}(qh)
-\frac{qh}{2}\operatorname{sgn}(\lambda_{R,\alpha})\right].
\label{eq:fourD_benchmark}
\end{equation}
The sum runs over the four pairs $(R,q)$ in Eq.~(\ref{eq:branch}), with the representation dimensions included in the eigenvalue sum over $\alpha$. The local cubic term cancels, but the finite part can be nonzero. Indeed, $\sum_q n_q q^5=240$, allowing a fifth-order contribution. Equation~(\ref{eq:fourD_benchmark}) therefore gives the chiral phase independently of the 5D APS spectrum. Local gauge-invariant counterterms depend on the curvature and its covariant derivatives~\cite{Dai:1994kq,Witten:2019bou}. For the backgrounds in Eq.~(\ref{eq:background_general}), these terms do not depend on $h$ and hence cancel in the ratio $W(h)/W(0)$.

For an $SU(4)$ connection, let $B=B_j\,dx^j$, with the matrices in each trace taken in representation $R$. We write $\eta_R(B)$ for $\eta(H_{3,R})$ at $a_j=0$ and $\eta_R(0)$ for its free-field value. Their difference is given locally by the 3D Chern--Simons form~\cite{Alvarez-Gaume:1984zst,Witten:2019bou},
\begin{equation}
\begin{aligned}
\eta_R(B)-\eta_R(0)={}&\frac{1}{4\pi^2}\int_{T^3}\operatorname{tr}_R\bigl[
B\wedge dB\\
&+\frac{2i}{3}B\wedge B\wedge B\bigr].
\end{aligned}
\label{eq:eta_nonabelian}
\end{equation}
We use this relation along deformations with no eigenvalue crossing, in the Hermitian convention specified above. The trace is proportional to the quadratic index of $R$. For the shifts relative to the free operator, $\eta_{\mathbf6}(B)-\eta_{\mathbf6}(0)=2[\eta_{\mathbf4}(B)-\eta_{\mathbf4}(0)]$ and the corresponding shift in $\overline{\mathbf4}$ equals that in $\mathbf4$. The $B^3$ term contains the antisymmetric structure constants of $SU(4)$, and its trace therefore has the same quadratic-index factor in the conjugate representation. The coefficients $2,-3,-1$ in the complete phase therefore cancel these local shifts. Thus, in the absence of spectral flow, the remaining non-Abelian phase is also given by the convergent sum in Eq.~(\ref{eq:fourD_benchmark}).

%%%%%%%%%%%%%%%%%%%%%%%%%%%
\subsection{The 5D APS phase and bulk extensions}
\label{sec:aps}

\begin{figure*}[t]
\centering
\includegraphics[width=0.80\textwidth]{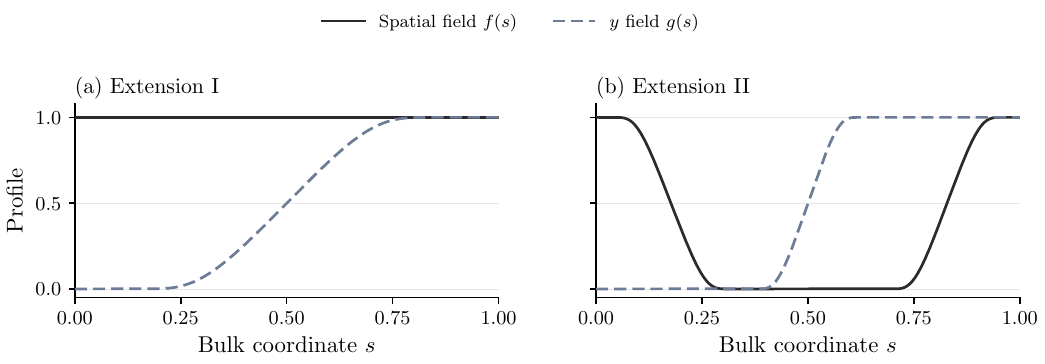}
\caption{Two smooth bulk extensions between the same boundary fields. The solid curves are the spatial-field coefficients $f(s)$ and the dashed curves are the coefficients $g(s)$ of $A_y$. Both profiles are constant near the two boundaries. Only extension II changes the spatial field in the interior.}
\label{fig:extensions}
\end{figure*}

For the 5D calculation, we use the cylinder
\begin{equation}
Y_5=T^4\times[0,1],\qquad s\in[0,1].
\label{eq:cylinder}
\end{equation}
The boundary at $s=0$ carries the reference field with $A_y=0$, and that at $s=1$ carries the target field with $A_y=hY$. Their induced orientations are opposite. We use a product metric and a smooth gauge connection that is independent of $s$ near either boundary. At the two boundaries we impose the APS spectral projectors of the corresponding four-dimensional Dirac operators~\cite{Atiyah:1975jf,Witten:2026ezr}. The projectors select complementary boundary eigenspaces because the outward normals point in opposite directions. This is the same type of bulk extension used in our two-dimensional calculation~\cite{Misumi:2026contf}, now with a five-dimensional operator.

We choose two smooth bulk extensions of the same boundary gauge fields. Both can be written as
\begin{equation}
 A_j(s)=f(s)A_j^{\mathrm{target}},\qquad A_y(s)=g(s)hY,
\label{eq:path_profiles}
\end{equation}
where $A_j^{\mathrm{target}}$ is the spatial gauge field at $s=1$ from Eq.~(\ref{eq:background_general}). The functions $f(s)$ and $g(s)$ are smooth and constant in collars near $s=0$ and $s=1$. Both profiles satisfy $f(0)=f(1)=1$, $g(0)=0$, and $g(1)=1$. For extension I, $f(s)=1$ and $g(s)$ increases from zero to one. For extension II, the spatial field is first turned off, then $g(s)$ increases, and finally the spatial field is restored. Figure~\ref{fig:extensions} shows both profiles.

We write the Hermitian 5D Dirac operator as $H_5=i\Gamma_s\partial_s+\mathcal B(s)$, where $\Gamma_s$ is the gamma matrix in the fifth direction and $\mathcal B(s)$ contains the 4D Dirac operator at fixed $s$. We impose the APS projectors at both ends and diagonalize the resulting Hermitian operator in a constrained Galerkin basis. For the signed eigenvalues $E_\ell$, we evaluate the heat-kernel-regularized spectral asymmetry
\begin{equation}
\eta_\delta(H_5)=\sum_\ell\operatorname{sgn}(E_\ell)
\operatorname{erfc}(\sqrt\delta|E_\ell|).
\label{eq:eta_heat}
\end{equation}
Here $\operatorname{erfc}(x)=(2/\sqrt\pi)\int_x^\infty e^{-t^2}dt$ is the complementary error function. With torus periods set to one, eigenvalues are measured in inverse-length units and $\delta$ has dimensions of length squared. We always quote $\delta$ together with the basis cutoffs. For each sector $(R,q)$, let $H_{5,R,q}$ denote the corresponding 5D operator. We sum its spectral asymmetry over the four sectors as $\eta_{15}(\delta)=\sum_{R,q}\eta_\delta(H_{5,R,q})$ and define the full phase by $\Theta_{\mathrm{APS}}(\delta)=\pi\eta_{15}(\delta)$. The orientation is chosen to match the Weyl block in Eq.~(\ref{eq:weyl}). Reversing it gives the $\exp(-i\pi\eta_Y)$ convention of Ref.~\cite{Witten:2026ezr}, where $\eta_Y$ denotes the 5D APS $\eta$-invariant on $Y_5$. The same sign is used for all charges and both extensions.

A finite matrix has an integer signed eigenvalue count in the limit $\delta\to0$. In our finite bases the count is zero. Thus the spectral cutoffs must be increased at fixed $\delta$ before extrapolating toward $\delta=0$. We denote by $N$ the Fourier cutoff in $x,w$, by $K_z$ the $z$ cutoff, by $K_y$ the antiperiodic $y$ cutoff, and by $L$ the polynomial basis size along $s$. Extension I can be reduced to a set of one-dimensional APS problems labeled by the eigenvalues of $H_3$. Extension II changes $H_3(s)$ and requires a direct five-dimensional matrix calculation. These methods and their independent checks are described in Appendix~\ref{app:numerics}.

%%%%%%%%%%%%%%%%%%%%%%%%%%%%%%%%%%%%%%%%%%%%%%%%%%%%%%%%%%%%%%%%%%%%
\section{Numerical results for Abelian backgrounds}
\label{sec:abelian}

%%%%%%%%%%%%%%%%%
\subsection{4D phase benchmark}

For the background in Eqs.~(\ref{eq:abelian_background}) and (\ref{eq:parameters}), the Fourier representation of $H_3$ separates into blocks labeled by the $x$ and $w$ momenta. The $z$ modes are coupled by the sine and cosine background. We diagonalize these blocks and evaluate Eq.~(\ref{eq:fourD_benchmark}) without a heat-kernel regulator. Table~\ref{tab:fourD} gives the cutoff dependence. The final value is
\begin{equation}
\Theta_{4D}=1.483354397942\times10^{-3}\ \mathrm{rad}.
\label{eq:target_phase}
\end{equation}

\begin{table}[t]
\begin{ruledtabular}
\begin{tabular}{ccc}
$N$ & $K_z$ & $\Theta_{4D}$ (rad)\\
\hline
2&4&0.001483357181671\\
3&6&0.001483354411451\\
4&8&0.001483354397986\\
5&8&0.001483354397942\\
\end{tabular}
\end{ruledtabular}
\caption{Four-dimensional chiral phase for the Abelian background, obtained from the convergent spectral sum. The spatial cutoff is $n_x,n_w=-N,\ldots,N$, and the $z$ cutoff is $|n_z|\le K_z$.}
\label{tab:fourD}
\end{table}

The free spatial Dirac operator has a gap $\pi\sqrt3$ with the antiperiodic spin structure. For the Abelian background, the gauge-field perturbation is bounded by $|q|\max(|u|,|v|)\le2.4$, which is smaller than this gap. Thus no eigenvalue crosses zero when the spatial field is turned on. Before multiplying by $n_q$, the phases for charges $q=-3$ and $q=4$ are $-0.14838$ and $+0.35411$ rad. Their combination with the other sectors leaves the much smaller but nonzero phase in Eq.~(\ref{eq:target_phase}). This is the finite remainder after cancellation of the cubic Chern--Simons terms. To check its weak-field behavior, we set $(u,v,h)=t(0.2,0.3,0.17)$. The phase increases from $2.40\times10^{-8}$ rad at $t=0.25$ to $8.03\times10^{-7}$ rad at $t=0.50$, a factor of $33.4$ compared with $2^5=32$. This supports the fifth-order behavior and is consistent with earlier studies of anomaly-free chiral determinants~\cite{Bornyakov:1999be}. The result in Eq.~(\ref{eq:target_phase}) comes from the full spectral sum, without a weak-field expansion.

%%%%%%%%%%%%%%%%%%%%%%%%%
\subsection{5D phases from two bulk extensions}

We next calculate the five-dimensional phase for the two bulk extensions in Fig.~\ref{fig:extensions}. Extension I uses a fixed spatial operator and separates into one-dimensional APS problems. Extension II changes the spatial operator in the fifth direction and is evaluated with a coupled five-dimensional matrix. The main numerical settings are $(N,K_z,K_y,L)=(9,9,7,28)$ and $(5,5,6,28)$, respectively. We compare the same values of the heat-kernel parameter $\delta$ for both extensions and extrapolate within $0.008\le\delta\le0.020$.

The extrapolation uses information from the short-time heat-kernel expansion. In the 5D bulk, parity-odd local terms of the dimensions relevant to the first two positive powers of $\delta$ are cubic in the gauge field strength, with charge dependence proportional to $q^3$. Power counting allows terms proportional to $q^5$ starting at order $\delta^3$. Since $\sum_q n_qq^3=0$, this suggests that the terms linear and quadratic in $\delta$ cancel in the complete representation. The charge-resolved numerical data support this expectation, as Table~\ref{tab:heat_slopes} shows. For a single charge-$q$ component, denote its 5D spectral asymmetry by $\eta_q(\delta)$, before multiplying by $n_q$. We write the normalized finite-difference slope as $S_q=[\eta_q(\delta_2)-\eta_q(\delta_1)]/[q^3(\delta_2-\delta_1)]$. Its weak dependence on $q$ shows the common cubic contribution. We fit these slopes to $S_q=\beta_0+q^2\beta_2$ on each interval. A contribution proportional to $q^5\delta^3$ in $\eta_q$ would give a $q^2\delta^2$ term in $S_q$. Dividing the fitted $\beta_2$ by the square of the interval midpoint gives about $0.023$ and $0.024$ for the two intervals. This is consistent with a fifth-power charge contribution beginning at order $\delta^3$ in $\eta_q$. We therefore fit the total spectral asymmetry with
\begin{equation}
\eta_{15}(\delta)=c_0+c_3\delta^3+c_4\delta^4.
\label{eq:fit}
\end{equation}
Here $c_0$ is the extrapolated value of $\eta_{15}$ at $\delta=0$, and $c_3,c_4$ are fit coefficients. This argument concerns the local bulk terms. APS boundary conditions can also produce nonlocal and logarithmic contributions to heat traces~\cite{Vassilevich:2003xt,Grubb:2003apslogs}. We have not proved that all such terms cancel in the present relative problem. For this reason, the charge-resolved slopes and the stability under changes of the fitting interval are part of the numerical evidence rather than a substitute for a proof.

\begin{table}[t]
\begin{ruledtabular}
\begin{tabular}{crr}
$q$ & $[0.010,0.012]$ & $[0.008,0.010]$\\
\hline
$2$ & $-0.0125083$ & $-0.0126819$\\
$-3$ & $-0.0124945$ & $-0.0126723$\\
$-1$ & $-0.0125166$ & $-0.0126876$\\
$4$ & $-0.0124751$ & $-0.0126588$\\
\end{tabular}
\end{ruledtabular}
\caption{Charge-resolved heat-kernel slopes for Abelian extension I, normalized by $q^3$. Each entry is $S_q$, defined in the text, with the intervals shown as column headings. The setting is $(N,K_z,K_y,L)=(9,9,6,28)$. The nearly common slopes test the cancellation of the low-order regulator dependence.}
\label{tab:heat_slopes}
\end{table}

\begin{figure}[t]
\centering
\includegraphics[width=0.93\columnwidth]{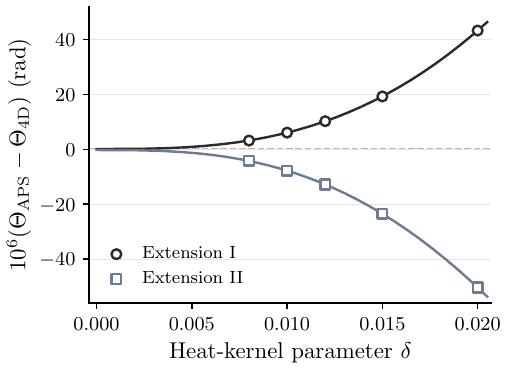}
\caption{Difference between the 5D APS phase and the independent 4D result for the Abelian background. Circles (squares) show extension I (II), and the solid curves are fits to Eq.~(\ref{eq:fit}) over $0.008\le\delta\le0.020$. The dashed horizontal line marks zero difference. The fitted intercepts are compared in Table~\ref{tab:two_paths}.}
\label{fig:abelian_aps}
\end{figure}

\begin{table*}[t]
\begin{ruledtabular}
\begin{tabular}{ccccc}
Fit interval & $\Theta^{\mathrm I}_{\mathrm{APS}}$ (rad) & $\Theta^{\mathrm{II}}_{\mathrm{APS}}$ (rad) & Difference from $\Theta_{4D}$ (I, II) & Difference between extensions\\
\hline
$[0.008,0.020]$ & $0.001483464222$ & $0.001483099038$ & $+0.0074\%,\ -0.0172\%$ & $0.0246\%$\\
$[0.010,0.020]$ & $0.001483521102$ & $0.001482965604$ & $+0.0112\%,\ -0.0262\%$ & $0.0374\%$\\
\end{tabular}
\end{ruledtabular}
\caption{Five-dimensional APS phases after extrapolation in $\delta$. Superscripts I and II label the two bulk extensions. The four-dimensional benchmark is $\Theta_{4D}=0.001483354397942$ rad. The percentage differences refer to this value. Extension I uses $(N,K_z,K_y,L)=(9,9,7,28)$ and extension II uses $(5,5,6,28)$.}
\label{tab:two_paths}
\end{table*}

Figure~\ref{fig:abelian_aps} shows the heat-kernel dependence for both bulk extensions. Their finite-$\delta$ behavior differs, yet both extrapolated phases agree with the independent 4D result. Table~\ref{tab:two_paths} shows that the largest difference from the 4D value is $0.0262\%$ for the two fitted intervals, while the two extensions differ by at most $0.0374\%$. Thus both extensions reproduce the 4D phase to better than $0.03\%$ in these fits. This is our central numerical result. In extension II, the spatial gauge field changes along the fifth direction, so the agreement directly tests a coupled 5D APS spectrum rather than only a fixed spatial operator.

We also increase the spectral cutoffs separately. Table~\ref{tab:directional} shows the change in the total $\eta$ of extension II at $\delta=0.008$. Of these variations, the $y$-direction cutoff has the largest effect, and the main results in Table~\ref{tab:two_paths} include the higher $K_y=6$ setting. The $N=6$ and $K_z=6$ tests were performed separately at $K_y=5$. These checks support the stability of the finite-basis calculation in the tested range.

\begin{table}[t]
\begin{ruledtabular}
\begin{tabular}{cc}
Cutoff change & Change in $\eta_{15}$\\
\hline
$N:5\to6$ & $+1.44\times10^{-11}$\\
$K_z:5\to6$ & $+1.47\times10^{-11}$\\
$K_y:5\to6$ & $-1.10\times10^{-7}$\\
\end{tabular}
\end{ruledtabular}
\caption{Changes in the total APS $\eta$ for Abelian extension II at $\delta=0.008$. We increase one cutoff at a time from $(N,K_z,K_y,L)=(5,5,5,28)$.}
\label{tab:directional}
\end{table}

%%%%%%%%%%%%%%%%%%%%%%
\subsection{Single-charge anomaly check}

We also calculate the difference in the 5D $\eta$-invariants of the two extensions for a single charge $q=4$. Here $A$ is the unit-charge Abelian gauge one-form on $Y_5$ and $F=dA$. We use $\Delta$ for extension I minus extension II and write $\eta_{q=4}^{\mathrm I}$ and $\eta_{q=4}^{\mathrm{II}}$ for the $\delta\to0$ APS invariants of this single-charge field in the two extensions. Their difference is fixed by the five-dimensional Abelian Chern--Simons term. In the orientation used above,
\begin{equation}
\Delta\!\int_{Y_5} A\wedge F\wedge F=-6\pi kuvh,
\label{eq:cs5difference}
\end{equation}
and the predicted difference in $\eta$ is
\begin{equation}
\eta_{q=4}^{\mathrm I}-\eta_{q=4}^{\mathrm {II}}
=\frac{4^3kuvh}{4\pi^2}
=0.132284937363.
\label{eq:cs4target}
\end{equation}
This is a separate test of the sign and normalization of the anomaly. At $(N,K_z,K_y,L)=(5,5,5,28)$, linear and quadratic extrapolations over $[0.008,0.020]$ give relative differences of $0.46\%$ and $0.11\%$ from Eq.~(\ref{eq:cs4target}), respectively. Over $[0.010,0.020]$, the corresponding differences are $0.55\%$ and $0.27\%$. Thus the quadratic extrapolations reproduce the Chern--Simons coefficient within $0.3\%$ in both fit intervals, while the linear fits agree within $0.6\%$. The sign is reproduced in every case.

%%%%%%%%%%%%%%%%%%%%%%%%%%%%%%%%%%%%%%%%%%%%%%%%%%%%%%%%%%%%%%%%%%%%%
\section{Numerical results for non-Abelian backgrounds}
\label{sec:nonabelian}

%%%%%%%%%%%%%%%%%%%%%%
\subsection{4D phases in non-Abelian fields}

For the non-Abelian tests, we set $a_j=0$ in Eq.~(\ref{eq:background_general}) and choose noncommuting constant matrices $B_j$. Their field strength is nonzero even though the potentials are coordinate independent,
\begin{equation}
F_{ij}=i[B_i,B_j].
\label{eq:nonab_curvature}
\end{equation}
Thus the constant potentials define a genuinely non-Abelian curvature. Equation~(\ref{eq:kk}) still applies because all $B_j$ commute with $A_y$ and are independent of $y$.

We first consider an $SU(2)$ subgroup acting on the first two colors, with $B_x=\tau_1$, $B_z=\tau_2$, and $B_w=\tau_3$ in that block. Here $\tau_1,\tau_2,\tau_3$ are the Pauli matrices, normalized by $\operatorname{tr}(\tau_i\tau_j)=2\delta_{ij}$. We use the Pauli matrices without a factor of $1/2$ in these gauge potentials. We then consider matrices that generate the full $\mathfrak{su}(4)$ Lie algebra. Define
\begin{equation}
X_{ij}=\frac{E_{ij}+E_{ji}}2,\qquad
Z_{ij}=\frac{-iE_{ij}+iE_{ji}}2,
\label{eq:generators}
\end{equation}
where $(E_{ij})_{ab}=\delta_{ia}\delta_{jb}$ are elementary $4\times4$ matrices, and choose
\begin{align}
B_x&=2X_{12}+1.3X_{34},\nonumber\\
B_z&=1.8X_{23}+Z_{14},\nonumber\\
B_w&=1.9Z_{13}+1.1X_{24}.
\label{eq:su4_B}
\end{align}
These matrices generate all 15 generators of $\mathfrak{su}(4)$, rather than an embedded $SU(2)$ subgroup. We construct the $\mathbf6$ as the antisymmetric product of two fundamentals and the $\overline{\mathbf4}$ as the conjugate representation. Both the 4D and 5D calculations include all 15 Weyl components.

\begin{table}[t]
\begin{ruledtabular}
\begin{tabular}{ccc}
$N$ & $SU(2)$ subgroup (rad) & Full $\mathfrak{su}(4)$ (rad)\\
\hline
2 & 0.001659985674 & 0.000930766517\\
3 & 0.001660031148 & 0.000930798723\\
4 & 0.001660031235 & 0.000930798785\\
5 & 0.001660031235 & 0.000930798785\\
\end{tabular}
\end{ruledtabular}
\caption{Four-dimensional chiral phase in the two non-Abelian backgrounds, evaluated with Eq.~(\ref{eq:fourD_benchmark}). The same half-open Fourier cutoff $n_j=-N,\ldots,N-1$ is used in $x,z,w$.}
\label{tab:nonab_fourD}
\end{table}

Table~\ref{tab:nonab_fourD} shows the Fourier-cutoff convergence. The full $SU(4)$ example gives
\begin{equation}
\Theta_{4D}^{SU(4)}=0.000930798785\ \mathrm{rad}.
\label{eq:nonab_target}
\end{equation}
Increasing $N$ from 4 to 5 changes the phase by about $10^{-13}$ rad. The phase is unchanged by a common constant $SU(4)$ gauge transformation of all three matrices. The free spatial Dirac operator has a gap $\pi\sqrt3\simeq5.44$. For the deformation $B_j\to tB_j$ with $0\le t\le1$, the perturbation norm is bounded by $\sum_j\lVert B_j^R\rVert$. For the full $SU(4)$ background these bounds are $2.85$ in $\mathbf4$ and $\overline{\mathbf4}$ and $4.55$ in $\mathbf6$. For the embedded $SU(2)$ background, the bound is $3.00$ in each nontrivial sector. All are below the free gap, so no spatial eigenvalue crosses zero along either deformation. For the full $SU(4)$ background, the phase changes sign when one of the three spatial matrices is reversed,
as expected for a parity-odd quantity. The local Chern--Simons formula gives
$\eta_{\mathbf4}(B)-\eta_{\mathbf4}(0)=-0.0587663$ for this background.
The corresponding shifts in $\mathbf6$ and $\overline{\mathbf4}$
are twice and equal to this value, respectively, in the branch without zero-mode crossings. Thus the non-Abelian
Chern--Simons contributions cancel by Eq.~(\ref{eq:mixed}), while the
finite spectral contribution gives the nonzero phase.

%%%%%%%%%%%%%%%%%%%%%%%%%%%%%
\subsection{5D APS phases}

We report the non-Abelian results at finite heat-kernel parameters, since the extrapolation form used for the
Abelian background in Sec.~\ref{sec:abelian} has not been established for non-Abelian fields.

%We first compute the 5D APS phase with the non-Abelian spatial field held fixed along the cylinder. 
We first compute the 5D APS phase for the full $SU(4)$ background using Extension I defined in Sec.~\ref{sec:aps}.
The APS operator then separates into sectors labeled by the eigenvalues of $H_3$. Because the spatial matrices are constant, Fourier modes in $x,z,w$ are conserved, and a single cutoff $N$ applies to all three directions. For the full $\mathfrak{su}(4)$ example, the APS result at $(N,K_y,L)=(4,6,28)$ differs from the independent 4D phase by $-0.113\%$, $-0.025\%$, $+0.014\%$, $+0.067\%$, $+0.198\%$, and $+0.493\%$ at $\delta=0.020$, $0.015$, $0.012$, $0.010$, $0.008$, and $0.006$, respectively. All six finite-regulator results agree with the 4D phase within $0.5\%$, and the three values at $\delta=0.010$--$0.015$ agree within $0.07\%$. A direct 5D matrix calculation and the reduced fixed-$H_3$ method also agree to better than $10^{-12}$ in $\eta$ at common low cutoffs.
These are independent checks of the spectral implementation. 
%The fixed spatial field makes this calculation separable in the eigenbasis of $H_3$. The Abelian extrapolation of Sec.~\ref{sec:abelian} uses charge-resolved information specific to that background. We report the non-Abelian results at finite heat-kernel parameters rather than assume the same extrapolation form.

\begin{figure}[t]
\includegraphics[width=0.93\columnwidth]{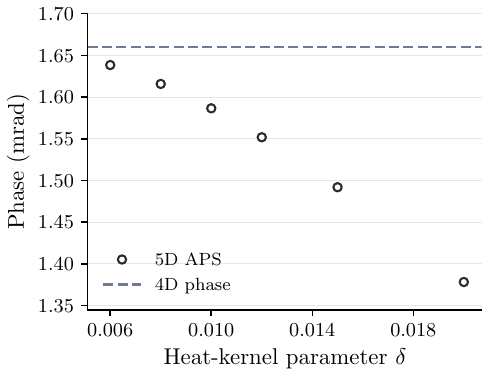}
\caption{5D APS phase for the nonseparable bulk extension with an embedded
$SU(2)$ background, compared with the independent 4D phase (dashed line).
The 5D calculation uses $(N,K_y,L)=(4,6,28)$. The plotted values are
at finite heat-kernel parameter $\delta$. No extrapolation is shown.}
\label{fig:nonab_aps}
\end{figure}

%For a nonseparable test, we use the embedded $SU(2)$ field, with the spatial gauge connection varying along the fifth direction.
For a nonseparable test, we use Extension II with the embedded $SU(2)$ background, where the spatial gauge connection varies along the fifth direction.
The 5D problem cannot then be separated into independent sectors labeled by the eigenvalues of a fixed $H_3$.
We use the smaller embedded $SU(2)$ background because a corresponding
calculation with the full $SU(4)$ field is substantially more expensive.
Figure~\ref{fig:nonab_aps} compares the resulting finite-$\delta$ phase
with its independently calculated 4D value at $(N,K_y,L)=(4,6,28)$.
The difference decreases to about $1.3\%$ at $\delta=0.006$.
At $(N,K_y)=(3,4)$ and $\delta=0.008$, increasing the polynomial
cutoff from $L=24$ to $L=28$ changes $\eta$ by about $6.5\times10^{-10}$.
The Fourier-cutoff changes are also smaller than the finite-$\delta$
difference near $\delta=0.008$. These checks do not establish convergence
at smaller $\delta$, and the extrapolated phase still depends on the
fit function. We therefore do not assign an extrapolated accuracy to this
nonseparable test.

The full $SU(4)$ example demonstrates close agreement of the 5D APS phase with the independent 4D result for a genuinely non-Abelian field. The embedded $SU(2)$ example further tests a nonseparable 5D bulk extension in which the spatial gauge field varies along the fifth direction.

%%%%%%%%%%%%%%%%%%%%%%%%%%%%%%%%%%%%%%%%%%%%%%%%%%%%%%%%%%%%%%%%%%%%%%%
\section{Summary and discussion}
\label{sec:summary}

We have numerically calculated 5D APS $\eta$-invariant phases for the anomaly-free 4D chiral fermion representation $\mathbf{10}\oplus\overline{\mathbf5}$ of $SU(5)$. For gauge fields independent of one torus coordinate, with a commuting constant component along that direction, the 4D chiral phase can be calculated separately from a convergent sum over 3D Dirac eigenvalues. This makes it possible to test the 5D APS phase after the local anomalous terms have canceled.

For the Abelian background, two different 5D bulk extensions reproduce the independent 4D phase to better than $0.03\%$ in the two tested extrapolation intervals. The extensions agree with each other to better than $0.04\%$. Most importantly, extension II varies the spatial gauge field along the fifth direction and requires a coupled 5D spectral calculation. We also reproduce the sign and normalization of the single-charge Chern--Simons coefficient to within $0.3\%$ using quadratic extrapolations.

The 5D APS calculation also agrees with the independently calculated phase for a non-Abelian field generating the full $\mathfrak{su}(4)$ algebra. The differences are within $0.5\%$ across all six tested heat-kernel parameters, with differences below $0.07\%$ at three intermediate values. 
We also study a nonseparable bulk extension with an embedded $SU(2)$ background, where the spatial gauge
field varies along the fifth direction. The finite-$\delta$ APS phase moves closer to the independent 4D result as $\delta$ decreases, although
a stable extrapolation to $\delta=0$ has not yet been obtained.

Our calculation isolates the 5D APS phase in smooth Abelian and non-Abelian backgrounds with $F_{y\mu}=0$ and $\operatorname{tr}F\wedge F=0$. It provides a phase test complementary to the full complex-weight calculations of Ref.~\cite{Misumi:2026contf}. Extending the calculation to general 4D gauge fields and to ensembles reconstructed from lattice links is an important next step.

While this work was being finalized, a second-order formulation of chiral fermions with local boundary conditions was proposed~\cite{Yonekura:2026local}. It would be interesting to test this formulation numerically using the independent 4D phase benchmark developed here.

%%%%%%%%%%%%%
\begin{acknowledgments}
This work was partially supported by JSPS KAKENHI Grant Nos. 23K03425 and 22H05118. OpenAI ChatGPT was used to assist with numerical code development, figure preparation, and language editing. The author is responsible for the reported calculations and their scientific interpretation.
\end{acknowledgments}

%%%%%%%%%%%%%
\section*{Data availability}
The numerical data and the codes used for the spectral calculations and figures will be made available as ancillary files with the arXiv submission.

\appendix

%%%%%%%%%%%%%%%%%%%%%%%%%%%%%%%%%%%%%%%%%%%%%%%%%%%%%%%%%%%%%%%%%%%%%%
\section{The compact-direction product and finite Fourier cutoffs}
\label{app:product}

For antiperiodic modes $\omega_n=2\pi(n+1/2)$, let $\xi$ be a real holonomy shift, equal to $qh$ in the main text. The symmetric product for one three-dimensional eigenvalue $\lambda$ obeys
\begin{equation}
\prod_{n\in\mathbb Z}
\frac{\lambda+i(\omega_n+\xi)}{\lambda+i\omega_n}
=\frac{\cosh[(\lambda+i\xi)/2]}{\cosh(\lambda/2)}.
\label{eq:prod_app}
\end{equation}
Its absolute-value square is Eq.~(\ref{eq:magnitude}). The product may be proved by pairing $n$ and $-n-1$, using the standard infinite product for $\cosh$, and fixing the value to one at $\xi=0$. The paired window is also natural because the holonomy shifts the antiperiodic momenta. We use the continuous phase of the resulting complex number. A principal-value arctangent without continuous branch tracking would give an incorrect result when $|\xi|$ crosses $\pi$.

At finite Fourier cutoff, the antiperiodic modes are paired as $n=-K_y,\ldots,K_y-1$. A sharp cutoff in the PV-subtracted magnitude can still introduce a spurious holonomy dependence, even with this symmetric pairing. The Fourier cutoff must therefore be increased at fixed PV mass before the regulator mass is taken large. For the independent 4D benchmark, the analytic sum over the $y$ modes avoids this artifact. Its control for general gauge fields requires further study.

%%%%%%%%%%%%%%%%%%%%%%%%%%%%%%%%%%%%%%%%%%%%%%%%%%%%%%%%%%%%%%%%%%%%%
\section{Chern--Simons term for the two bulk extensions}
\label{app:cs}

For the Abelian background, let $A$ be the unit-charge $U(1)$ gauge one-form with $F=dA$. Write $A_x=uf(s)\cos(2\pi kz)$,
$A_w=vf(s)\sin(2\pi kz)$, and $A_y=hg(s)$. We use the orientation
$ds\wedge dx\wedge dy\wedge dz\wedge dw$. The Chern--Simons integral is
\begin{equation}
\int_{Y_5}A\wedge F\wedge F=
2\pi ku v h\int_0^1ds\,[2f'f g-2f^2g'].
\label{eq:cs5integral}
\end{equation}
Here primes denote derivatives with respect to $s$. For extension I, $f=1$ and $g$ increases from zero to one. Equation~(\ref{eq:cs5integral}) gives $-4\pi ku v h$. For extension II, $f$ decreases from one to zero before $g$ changes, and is restored after $g$ becomes one. The result is $+2\pi ku v h$. Their difference is therefore $-6\pi ku v h$, independent of the detailed shapes of the smooth switching functions. Using the anomaly coefficient $-q^3/(24\pi^2)$ for the phase and the relation $\Theta_{\mathrm{APS}}=\pi\eta$, we reproduce Eq.~(\ref{eq:cs4target}). The complete representation has $\sum_q n_qq^3=0$ and hence no local bulk-extension dependence from this term.

%%%%%%%%%%%%%%%%%%%%%%%%%%%%%%%%%%%%%%%%%%%%%%%%%%%%%%%%%%%%%%%%%%%%
\section{Spectral calculation and extrapolation}
\label{app:numerics}

For the Abelian background the spatial Fourier modes $n_x,n_w$ are conserved. We retain $-N\le n_x,n_w\le N$, $|n_z|\le K_z$, and $-K_y\le n_y<K_y$. In the spatial operator, the $\cos(2\pi kz)$ and $\sin(2\pi kz)$ terms couple neighboring $z$ modes. The reduced fixed-field calculation diagonalizes $H_3$ and then solves the one-dimensional APS boundary problem for each $\lambda$ and $n_y$. The nonseparable calculation retains the coupled $z$ modes through the bulk and diagonalizes the first-order five-dimensional operator in the APS-constrained basis.

The $s$ dependence is expanded in Legendre polynomials. We impose the negative/positive APS boundary subspaces at the opposite ends of the cylinder by a constraint matrix. For each momentum block the operator can be written as
\begin{equation}
H_5=i\rho_3\partial_s-\rho_2 H_3(s)-\rho_1 m_y(s),
\label{eq:5d_block}
\end{equation}
Here $\rho_i$ are Pauli matrices acting on an auxiliary two-component space, distinct from the spatial spin matrices in Eq.~(\ref{eq:H3}). Also, $m_y(s)=2\pi(n_y+1/2)+qh g(s)$ and $H_3(s)$ contains the spatial gauge field with profile $f(s)$. The tangential boundary operator is $\rho_1H_3(s)-\rho_2m_y(s)$. The APS constraint retains its negative-eigenvalue subspace at $s=0$ and positive-eigenvalue subspace at $s=1$, with the orientation of Sec.~\ref{sec:aps}. In an APS-constrained basis $\phi_a$, we construct
\begin{equation}
M_{ab}=\langle\phi_a,H_5\phi_b\rangle,\qquad
G_{ab}=\langle\phi_a,\phi_b\rangle.
\label{eq:galerkin}
\end{equation}
Here $M$ and $G$ are the projected operator and overlap matrices. We solve $M\mathbf c=EG\mathbf c$ for the Galerkin coefficient vector $\mathbf c$. This generalized Hermitian eigenvalue problem yields the signed eigenvalues entering Eq.~(\ref{eq:eta_heat}). For fixed spatial fields, a direct coupled-matrix calculation agrees with the reduced calculation to about $10^{-12}$ in $\eta$. For a test with spatially constant gauge potentials, the lowest eigenvalues also agree with the one-dimensional APS quantization equation to about $10^{-14}$ in the test cases. These checks validate the finite-matrix implementation, not the removal of the heat-kernel regulator.

We increase the Fourier and polynomial cutoffs at fixed $\delta$ before extrapolating toward $\delta=0$. For the complete chiral multiplet in the Abelian background, the measured charge dependence and the cancellation in Eq.~(\ref{eq:cancel}) motivate the fit $c_0+c_3\delta^3+c_4\delta^4$ in Eq.~(\ref{eq:fit}). We compare two fit windows and independently vary the spectral cutoffs. Possible contributions from APS boundary terms remain a limitation of the extrapolation ansatz, as discussed in Sec.~\ref{sec:abelian}. The non-Abelian varying-field results are given at finite $\delta$.

\bibliographystyle{utphys}
\bibliography{references}
\end{document}